\documentclass[useAMS,usenatbib]{mn2e}
\usepackage{graphicx}
\usepackage{times}
\usepackage{amssymb}
\usepackage{natbib}

\usepackage{pgf}
\usepackage{bm}

\def\gsim { \lower .75ex \hbox{$\sim$} \llap{\raise .27ex \hbox{$>$}} }
\def\lsim { \lower .75ex \hbox{$\sim$} \llap{\raise .27ex \hbox{$<$}} }

\begin{document}

\title[Satellites and Halos of Dwarf Galaxies]{Satellites and Halos of
Dwarf Galaxies}

\author[Sales et al.]{
\parbox[t]{\textwidth}{
Laura V. Sales$^{1}$,     
Wenting Wang$^{1,2}$,
Simon D. M. White$^{1}$ and
Julio F. Navarro$^{3}$
}
\\
\\
$^{1}$ Max-Planck Institute for Astrophysics, Karl-Schwarzschild-Strasse 1, 85740 Garching, Germany\\
$^{2}$ Key Laboratory for Research in Galaxies and Cosmology of
Chinese Academy of Sciences, Max-Planck-Institute Partner Group, \\
Shanghai Astronomical Observatory, Nandan Road 80, Shanghai 200030,
China\\
$^{3}$ Department of Physics and Astronomy, University of Victoria,
Victoria, BC, V8P 5C2, Canada
}
\maketitle

\begin{abstract} 

  We study the abundance of satellite galaxies as a function of
  primary stellar mass using the SDSS/DR7 spectroscopic catalogue. In
  contrast with previous studies, which focussed mainly on bright
  primaries, our central galaxies span a wide range of stellar mass,
  $10^{7.5} \leq M_*^{\rm pri}/M_\odot \leq 10^{11}$, from dwarfs to
  central cluster galaxies. Our analysis confirms that the average
  number of satellites around bright primaries, when expressed in
  terms of satellite-to-primary stellar mass ratio ($m_*^{\rm
    sat}/M_*^{\rm pri}$), is a strong function of $M_*^{\rm pri}$. On
  the other hand, satellite abundance is largely independent of
  primary mass for dwarf primaries ($M_*^{\rm pri}<10^{10} \,
  M_\odot$). These results are consistent with galaxy formation models
  in the $\Lambda$CDM scenario. We find excellent agreement between
  SDSS data and semi-analytic mock galaxy catalogues constructed from
  the Millennium-II Simulation.  Satellite galaxies trace dark matter
  substructure in $\Lambda$CDM, so satellite abundance reflects the
  dependence on halo mass, $M_{200}$, of both substructure and galaxy
  stellar mass ($M_*$). Since dark matter substructure is almost
  scale-free, the dependence of satellite abundance on primary mass
  results solely from the well-defined characteristic mass in the
  galaxy mass-halo mass relation. On dwarf galaxy scales, where models
  predict a power-law scaling, $M_* \propto M_{200}^{2.5}$, similarity
  is preserved and satellite abundance is independent of primary
  mass. For primaries brighter than the characteristic mass of the
  $M_*$-$M_{200}$ relation, satellite abundance increases strongly
  with primary mass. Our results provide strong support for the steep,
  approximately power-law dependence of dwarf galaxy mass on halo mass
  envisioned in $\Lambda$CDM galaxy formation models.
\end{abstract}


\section{Introduction}
\label{sec:intro}

Matching the galaxy luminosity function in the $\Lambda$CDM scenario
requires that the stellar mass of galaxies, $M_*$, should vary
strongly with the virial\footnote{Virial quantities are defined at the
  radius from the center of each halo where the mean enclosed density
  equals $200$ times the critical density of the Universe and are
  identified by a $200$ subscript. Units assume a Hubble constant of
  $H_0=73$ km s$^{-1}$ Mpc$^{-1}$ unless otherwise specified.} mass,
$M_{200}$, of their surrounding dark matter halos. This exercise
implies that the ``efficiency'' of galaxy formation, as measured by
the ratio $M_*/M_{200}$, decreases steadily toward both smaller and
larger masses from a maximum at $M_{200} \sim 10^{12} M_\odot$
\citep{Yang2003,Vale2004,Shankar2006,Zheng2007,Kravtsov2010,Moster2010,Behroozi2010,Guo2010,Guo2011_sam}. On
the scale of dwarf galaxies ($M_* < 10^{10}\, M_\odot$) these models
require a near power-law dependence, $M_* \propto M_{200}^{2.5}$, in
order to reproduce observations of faint objects \citep[however, see
][]{Behroozi2012}. Such a steep $M_*$-$M_{200}$ relation implies that
dwarfs differing by as much as three decades in stellar mass (e.g.,
from the Fornax dwarf spheroidal to the Large Magellanic Cloud)
inhabit halos spanning just over one decade in virial
mass. Furthermore, very few galaxies exceeding $10^6 \, M_\odot$ are
expected to have halos with virial masses below $10^{10}\, M_\odot$
\citep{Ferrero2012}.

These predictions have been recently challenged by a series of
observations, including (i) the lack of a characteristic velocity at
the faint-mass end of blind HI surveys (expected if most dwarfs live
in similar halos, \citealt{Zwaan2010,Papastergis2011}); and (ii) the
low virial mass (substantially below $10^{10}\, M_\odot$) inferred
from dynamical data for the dwarf spheroidal companions of the Milky
Way \citep{Boylan-Kolchin2012} and for nearby dwarf irregulars
\citep{Ferrero2012}. The evidence, however, is indirect, since halo
masses are estimated by extrapolating data that probe only the 
inner few kiloparsecs, where most baryons reside. 

We explore here the possibility of using satellite galaxies to help
constrain the virial masses of dwarf galaxies. The orbital motions of
satellite companions have often been used to estimate halo masses
\citep[see,
e.g.,][]{Zaritsky1997,Erickson1999,McKay2002,Prada2003,vandenBosch2004,Brainerd2005,Conroy2007,Wojtak2012},
but this work has largely been restricted to systems similar to or
brighter than the Milky Way. This is partly due to the difficulties in
obtaining redshifts for faint objects.  In addition, satellite
companions are less common around dwarf galaxies than around larger
systems: the number of satellites brighter than a certain fraction of
the primary luminosity, $N(>L^{\rm sat}/L^{\rm pri})$, declines strongly
toward fainter primaries \citep[e.g. ][]{GuoQuan2011,
  Wang_White2012}. Dwarf galaxy associations do exist, but only a
handful have been observed \citep[e.g.,][]{Tully2006}.

In $\Lambda$CDM, where satellite galaxies are thought to trace the
substructure of cold dark matter halos, satellite systems
are expected around all central galaxies, regardless of
luminosity. The number of satellites, and their dependence on primary
mass, should just reflect the abundance of substructure modulated by
the dependence of galaxy formation efficiency on halo mass.

Substructure abundance has been studied extensively through numerical
simulations, and shown to be nearly invariant with halo mass when
expressed as a function of satellite mass normalized to that of the
host \citep{Moore1999b,kravtsov2004,Gao2004,Wang2012}. This result,
together with the strict constraints on galaxy formation efficiency
mentioned above, imply that satellite number counts provide useful
information on the halo mass of dwarf galaxies. In particular, the
near self-similarity of cold dark matter halos provides an instructive
test: if satellite galaxies trace substructure then the abundance of
{\it luminous} satellites should also be scale-free on scales where
galaxy mass and halo mass are related by a power law.

We explore these issues here by identifying primary-satellite systems
in galaxy catalogues constructed from the Sloan Digital Sky Survey and
by comparing them with predictions from a semianalytic mock galaxy
catalogue based on the Millennium Simulations. The plan for this paper
is as follows. Sec.~\ref{sec:data} describes briefly the observational
and model datasets while Sec.~\ref{sec:results} presents our main
results. We summarize our main conclusions in Sec.~\ref{sec:concl}.

\section{Data and Catalogues}
\label{sec:data}

\subsection{Satellite and primary galaxies in SDSS/DR7}
\label{ssec:sdss}

We select primary galaxies spanning a wide range of stellar mass,
$7.5\leq \rm log(M_*/M_\odot) \leq 11$, from the spectroscopic New
York University Value Added Galaxy Catalog (NYU-VAGC).  This catalogue
was built on the basis of the seventh data release of the Sloan
Digital Sky Survey \citep[SDSS/DR7;][]{Blanton2005,Abazajian2009}.
 Stellar masses are taken directly from the NYU-VAGC catalog and
  are estimated by fitting stellar population synthesis models to the
  $k$-corrected galaxy colours. They assume a \citet{Chabrier2003}
  initial mass function.

We ensure isolation by imposing two conditions: (i) each primary must
be the brightest of all objects projected within its virial radius
with line-of-sight velocities differing by less than three times the
corresponding virial velocity; and (ii) no primary can be located
within the virial radius of a more massive system. Virial quantities
are inferred from the stellar mass, assuming the abundance-matching $M_*$-$M_{200}$
relation of \citet{Guo2010} and cosmological parameters consistent
  with WMAP1 results $\Omega_\lambda=0.75$, $\Omega_m=0.25$ and $H_0=73$
  km/s \citep{Spergel2003}. Fainter galaxies within a projected
distance $r_p < r_{\rm 200}$ and a line-of-sight velocity difference
$|\Delta V_{\rm l.o.s}|<3V_{200}$ are then classified as satellites
(see Sec.~\ref{ssec:semi-analytic}). We have checked that our results
are not sensitive to variations by factors of a few in these
thresholds, nor to the addition of $0.2$-$0.35$ dex scatter to the
$M_*$-$M_{200}$ relation.

We apply volume and edge corrections to our sample in the same way as
\citet{Wang_White2012}.  Completeness for SDSS spectroscopic data is
estimated to be $\sim 90\%$ for apparent $r$-band magnitudes brighter
than $m_r=17.7$ \citep{Blanton2005}. This limit applies to
  satellite and primary galaxies independently. However, satellites are
  fainter than their centrals, having an absolute magnitude difference
  $\Delta M=M_r^{\rm sat}-M_r^{\rm pri}$. This means that the
  effective volume where satellites of a given $\Delta M$ are (almost)
  complete varies strongly with $M_r^{\rm pri}$, or equivalently, with
  primary mass. For instance, satellites about a magnitude fainter than
  their hosts will have $M_r^{\rm sat} \sim -20.5$ and $-15.2$ for
  primaries with stellar masses in the ranges
  $10.5<\log(M_*/M_\odot)<11$ and $7.5<\log(M_*/M_\odot)<8$,
  respectively. In our sample, the limiting distance where satellites can 
  be found with $\Delta M =1$ is roughly $320$ kpc for primaries with
  $10.5<\log(M_*/M_\odot)<11$. This halves for primaries with 
  $M_*\sim10^{10} M_\odot$, and drops to $35$ $\rm Mpc$ for centrals with
  $7.5<\log(M_*/M_\odot)<8$. Despite the smaller volume surveyed for
  low mass galaxies, there are still enough galaxies to probe the satellite
  population of even the faintest primaries we considered.

One might think that the detection of faint objects would be aided by
selecting satellites from the photometric catalogue; which is complete
down to $\sim 4$ magnitudes fainter than the spectroscopic
sample. This requires the stacking of primary galaxies and a
statistical background subtraction in order to identify the excess
count corresponding to satellites
\citep{Lorrimer1994,Lares2011,Liu2011,Nierenberg2011,GuoQuan2011,Strigari2012,Wang_White2012}.
However, the signal-to-noise to carry out this subtraction is too low
for our faint primaries ($M_*^{\rm pri}<10^9 M_\odot$). Our sample is
then selected purely from the spectroscopic catalogue. We discuss
briefly possible biases affecting faint, low surface brightness
companions in Sec~\ref{sec:results}.

%%%%%%%%%%%%%%%%%%%%%%%%%%
\begin{center} \begin{figure} 
\includegraphics[width=84mm]{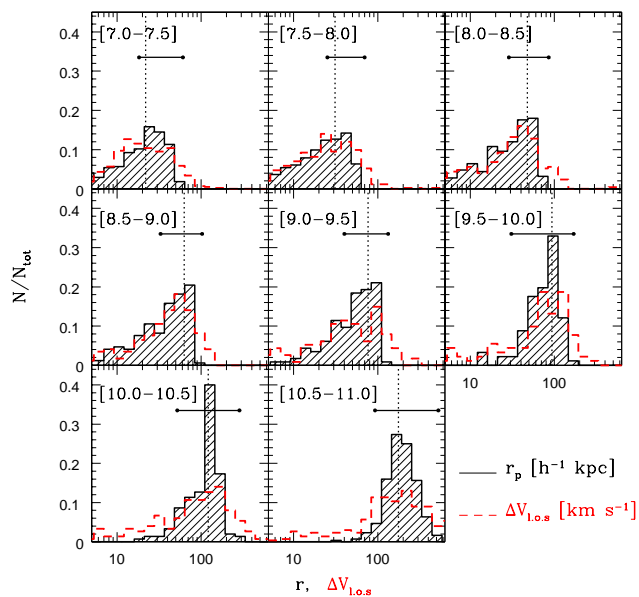} 
\caption{Distributions of projected distance (black) and line-of-sight
  velocity difference (red) for satellite galaxies in the semianalytic
  catalogue, binned by primary stellar mass. The minimum and maximum
  mass of each bin is quoted in each panel (in units of $\log(M_*^{\rm
    pri}/M_\odot)$).  The distributions peak at around the 
    average virial radius and velocity corresponding to a given bin
  (vertical dotted line). The minimum and maximum virial radius
    and virial velocity of the host haloes contributing to each bin
    is indicated by the horizontal line.  Notice that, by
    definition, satellites are located within the virial radius of
    their hosts. Their relative velocity, however, can exceed the
    virial value (see text for more details).}
\label{fig:deltas}
\end{figure}
\end{center}
%%%%%%%%%%%%%%%%%%

\subsection{Satellite and primary galaxies in the semi-analytic model}
\label{ssec:semi-analytic}

We compare our SDSS results with the semi-analytic catalogue of
\citet{Guo2011_sam}, based on the dark matter only Millennium-II Simulation
\citep[hereafter MS-II,][]{Boylan-Kolchin2010}. The model parameters
are carefully tuned to reproduce the observed abundance of
low-redshift galaxies over five orders of magnitude in stellar mass
and 9 mag in luminosity.
The semi-analytic data contain full 3D velocity and positional
information for all galaxies, and thus enables the evaluation of
potential biases that may be induced by the limited (projected) data
available in observational surveys.

Fig. 1 and 9 of \citet{Guo2011_sam} shows that predictions for galaxy
mass are reliable in simulated halos resolved with at least $\sim
200$-$300$ dark matter particles. This corresponds roughly to
$M_{200}\sim 2\times 10^9 M_\odot$ and $M_* \sim 10^6 M_\odot$ in
MS-II. We therefore consider only galaxies with $M_* \geq 10^6
M_\odot$ in the analysis below.

Galaxies in the semi-analytic catalogue inhabit dark matter halos and
subhalos identified using the {\sc subfind} group-finder algorithm
\citep{Springel2001a}. Primary galaxies are the central objects of
each halo; all other galaxies within the virial radius are considered
satellites. The catalogue also includes ``orphan'' galaxies whose
subhalos have been disrupted due to numerical resolution effects.
The catalogue contains
more than $157,000$ primary galaxies in the mass range $7 \leq \rm
log(M_*/M_\odot) \leq 12$ .

The semi-analytic data can be ``projected'' to mimic the same
satellite identification algorithm used for SDSS data (see, e.g.,
\citealt{Wang_White2012} for details).  Notice that, because of the
different identification criteria applied, the projected satellite and primary
samples selected from the mock catalogue differ from the 3D
samples (where we use information about the condition as
central/satellite object from the {\sc subfind} catalogues).  
This enables us to calibrate the parameters of the
identification procedure in order to minimize the contribution of
foreground and background objects in our primary/satellite sample.

Fig~\ref{fig:deltas} show the distributions of projected distance
($r_p$, shown in black) and line-of-sight velocity difference ($\Delta
V_{\rm l.o.s}$, shown in red) between primaries and ``true''
satellites, grouped in several bins of primary stellar mass. 
  Primaries in each bin span a range of virial masses given by the
  $M_*$-$M_{200}$ relation and its scatter. For each panel, a
  horizontal line indicates the minimum and maximum virial
  radius/velocity of the host haloes contributing to that bin. All
histograms peak approximately at the average virial radius and virial
velocity of host halos in the subsample, indicated by the vertical
dotted line.

Although by definition $r_p<r_p^{\rm max}=r_{200}$, the upper bound of the
velocity difference is less clear, as the escape velocity typically
exceeds the virial velocity of a halo substantially in the inner
regions. The red histograms in Fig.~\ref{fig:deltas} suggest that the
large majority of true satellites have line-of-sight velocities that
differ from their primaries by less than $\sim 3 V_{200}$. These
considerations justify the choices $r_p^{\rm max}=r_{200}$ and $\Delta
V_{\rm l.o.s}^{\rm max}=3\, V_{200}$ made to identify
satellite/primary systems in the observational sample
(Sec.~\ref{ssec:sdss}).

%%%%%%%%%%%%%%%%%%%%%%%%%%
\begin{center} \begin{figure} 
\includegraphics[width=84mm]{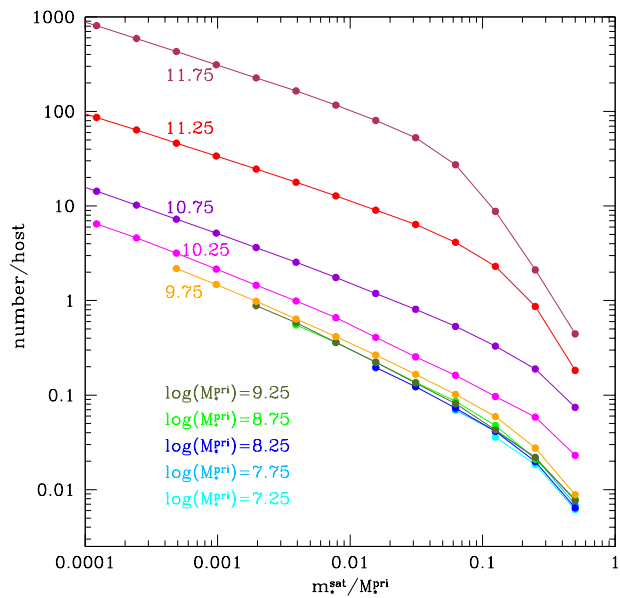} 
\caption{Cumulative number of satellite galaxies within the virial
  radius of primaries in the semi-analytic model of galaxy formation
  from \citet{Guo2011_sam}.  Symbols of different colors indicate the
  average number of satellites computed after binning primaries by
  mass in $0.5$ dex-width bins of $M_{\rm pri}$.  Note that, on
  average, the number of satellites decreases with decreasing primary
  mass down to $M_{\rm pri}\sim 10^{10}\, M_\odot$. Below this mass
  the scaled satellite mass function becomes independent of primary
  mass.}
\label{fig:mf_sam}
\end{figure}
\end{center}
%%%%%%%%%%%%%%%%%%

\section{Results}
\label{sec:results}

As discussed in Sec.~\ref{sec:intro}, the galaxy-halo mass relation is
expected to leave a clear imprint on the abundance of satellites
galaxies as a function of primary stellar mass. We explore these ideas
in Fig.~\ref{fig:mf_sam} using the semi-analytic catalogue described in
Sec.~\ref{ssec:semi-analytic}. 

Fig.~\ref{fig:mf_sam} shows, as a function of satellite-to-primary
mass ratio, the average number of satellites orbiting primaries of
different mass in the semi-analytic galaxy catalogue. Primaries are
binned in logarithmic $M_*$ bins of $0.5$ dex width; the central mass
value is quoted in the legend. Satellites are identified in 3D, using
the full position and velocity information available in the catalogue.

Fig. ~\ref{fig:mf_sam} shows clearly that the average number of
satellites of bright primaries
increases strongly with $M_*^{\rm pri}$.  On average, a primary as
massive as $10^{11.5} M_\odot$ is surrounded by roughly $10$
satellites more massive than $0.1\, M_*^{\rm pri}$. On the other hand,
only $\sim 40\%$ of primaries as massive as the Milky Way ($10^{10.75}
M_\odot$) have one satellite proportionally as massive. The
probability of having a companion with $m_*^{\rm sat}/M_*^{\rm
  pri}=0.1$ drops further to $\sim 10\%$ for $M_*^{\rm pri}=10^{10}
M_\odot$.

Interestingly, this trend does not hold for lower primary masses. The
satellite abundance, expressed in terms of $m_*^{\rm sat}/M_*^{\rm
  pri}$, becomes {\it independent of primary mass} in the dwarf-galaxy
regime ($M_*^{\rm pri} < 10^{10} M_\odot$).  As discussed in
Sec.~\ref{sec:intro}, this reflects the featureless power-law scaling
between galaxy and halo masses in these scales and is a prime
prediction of the model testable by observation.

%%%%%%%%%%%%%%%%%%%%%%%%%%
\begin{center} \begin{figure*} 
%%%
\includegraphics[width=0.475\linewidth]{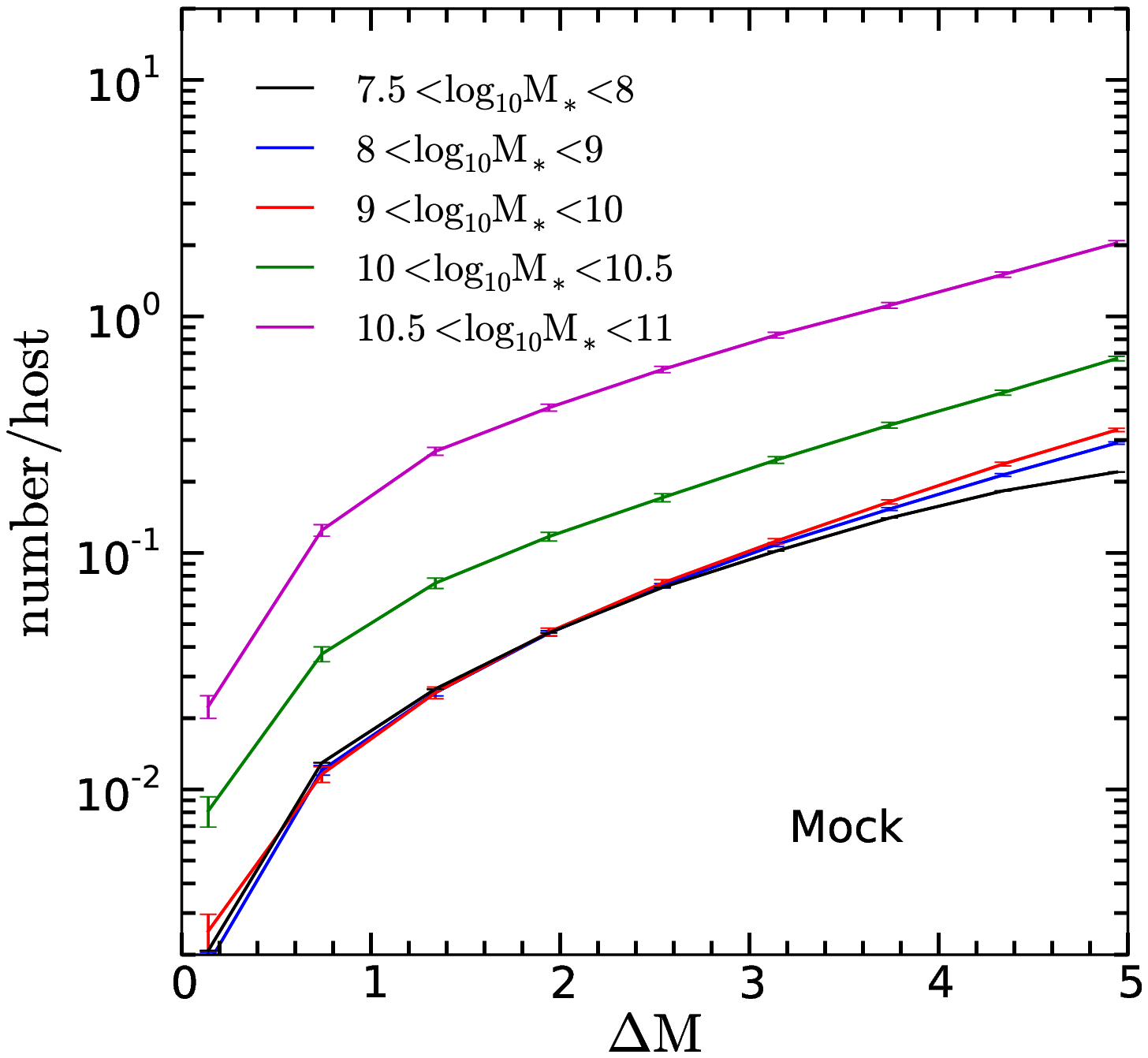} 
\includegraphics[width=0.475\linewidth]{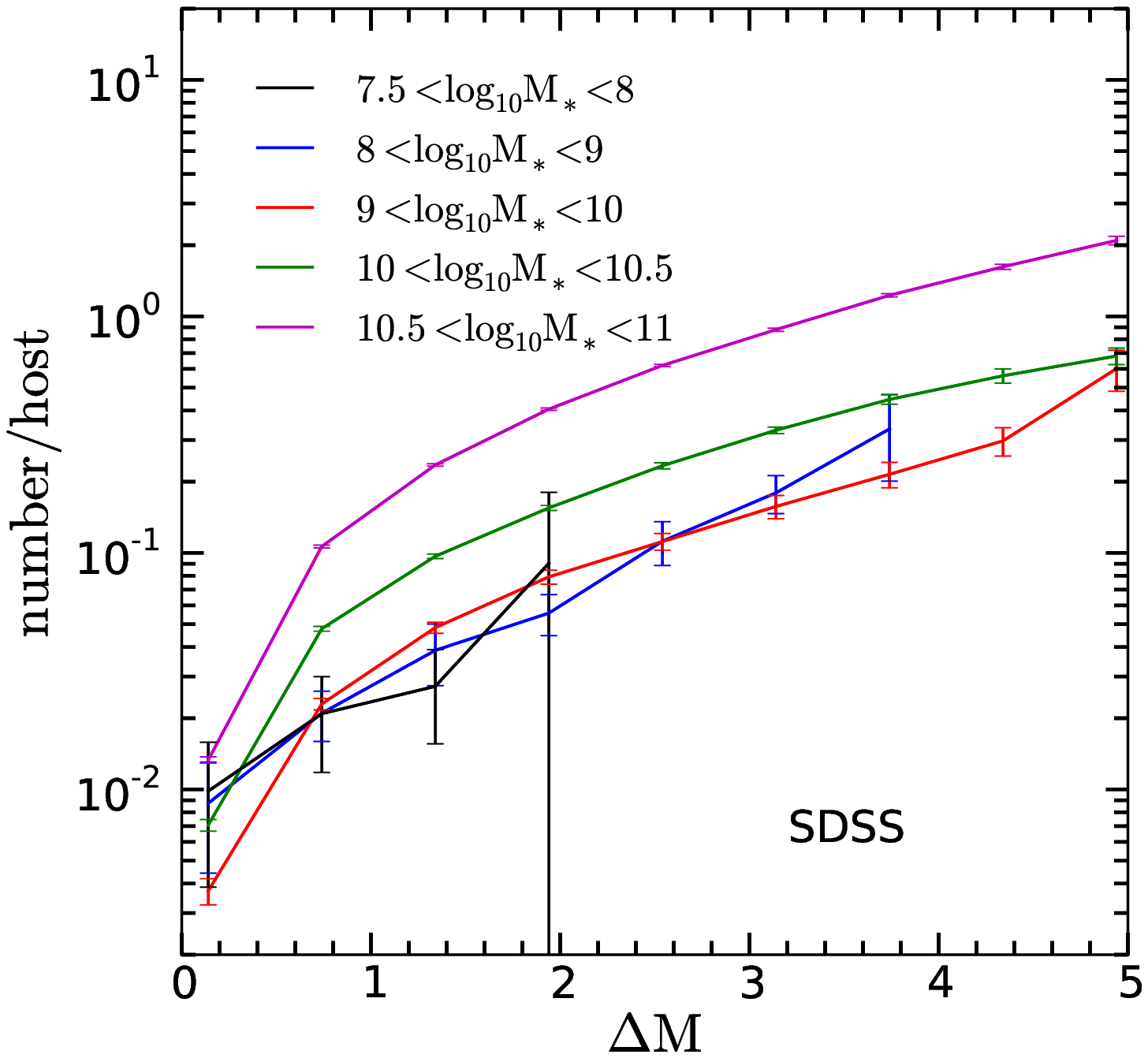} 
\caption{Average number of satellites per primary with a $r$-band magnitude
  difference smaller or equal to $\Delta M=M_r^{\rm sat}-M_r^{\rm
    pri}$. Each curve correspond to a given primary stellar mass
  range, as indicated by the labels. The left panel is for mock data
  from the semi-analytic catalogue. Primary/satellite galaxies are
  identified in projection, as outlined in Sec~\ref{ssec:sdss}. Error
  bars show uncertainties from 100 bootstrap resamplings. Note the
  lack of dependence on $M_*^{\rm pri}$ for $M_*^{\rm pri}<10^{10}
  M_\odot$.  {\it Right:} Same but for galaxies in the SDSS/DR7
  spectroscopic catalogue. As in the left panel, isolated dwarfs with
  stellar mass $7.5<\rm log(M_*^{pri}/M_\odot)<10$ seem to populate
  similar dark matter halos, $M_{\rm 200}\sim 10^{10}-10^{11} M_\odot$
  according to the simulations.}
\label{fig:lumfunc_sam_sdss}
\end{figure*}
\end{center}
%%%%%%%%%%%%%%%%%%

In order to take into account how projection effects and the
presence of interlopers may affect this result, we repeat the
analysis using only projected positions and line-of-sight velocities,
as described in Sec.~\ref{ssec:semi-analytic}. This enables us to
select primaries and satellites in identical ways for both model and
observational datasets.

Fig.~\ref{fig:lumfunc_sam_sdss} shows, for the mock (left) and SDSS
(right) samples, the average number of satellites with $r$-band
magnitude difference equal to or smaller than $\Delta M = M_r^{\rm
  pri} - M_r^{\rm sat}$. This is not only the simplest
  observational analogue of the stellar mass ratio but also allows a
  more straightforward derivation of the completeness correction for
  the catalogue.  The use of stellar masses would introduce extra
  uncertainties that would have to be compensated by more conservative
  cuts to ensure completeness. Working with magnitudes allows us to
  increase the effective statistical power of the catalogue. As in
Fig.~\ref{fig:mf_sam}, we bin primaries according to their stellar
mass, as indicated in the legends. Error bars correspond to 100
bootstrap resamplings of the data.

The left panel in Fig. ~\ref{fig:lumfunc_sam_sdss} shows that the
projected data behave similarly to the 3D sample: the abundance of
satellites at given $\Delta M$ increases with $M_*^{\rm pri}$ for bright
primaries \citep{GuoQuan2011,Wang_White2012} but becomes independent
of mass for dwarf primaries ($M_*^{\rm pri}<10^{10} M_\odot$).  The
overall behaviour is in strikingly good agreement with the SDSS
satellite abundances, shown on the right. Despite the large error bars
in the faint-primary bins (an unavoidable consequence of the limited
effective volume surveyed by SDSS) the observed trends in satellite
count with primary stellar mass closely resemble those in the mock
catalogue.  We interpret this result as providing strong evidence in
support of the nearly power-law dependence of galaxy mass on halo mass
on dwarf galaxy scales advocated by semianalytic models of galaxy
formation in the $\Lambda$CDM scenario.

One concern regarding this interpretation arises from the completeness
of SDSS spectroscopic data on dwarf galaxy scales. The sample is, on
average, $90\%$ complete at our apparent magnitude limit of
$m_r=17.7$. We partially correct for biases by weighting satellites and
primaries with their FGOTMAIN value, which characterizes the 
completeness of SDSS in a given region of the sky due to fiber 
collision \citep{Blanton2005}. However, the completeness might worsen 
if objects of low surface brightness, such as many dwarfs, are 
systematically missed. The results in Fig~\ref{fig:lumfunc_sam_sdss} 
would then represent lower-limits to the true satellite abundance, 
but would still provide useful constraints on the halo mass of their 
hosts. We notice, however, that if our results were strongly affected 
by low-surface brightness biases, the good agreement between 
observations and the semianalytic model would be puzzling. Nevertheless, 
this point requires further validation once surveys with improved 
surface brightness sensitivity become available.

 The deblending of extended objects into multiple spurious
  ``galaxies'' could potentially affect our results. As a sanity
  check, we have repeated the analysis of
  Fig.~\ref{fig:lumfunc_sam_sdss} using the NASA Sloan
  Atlas\footnote{http://www.nsatlas.org} (NSA). This catalogue is
  based on SDSS-DR8 data but uses a different background substraction
  technique that specifically improves the identification of galaxies
  over the original SDSS pipeline, especially for low-redshift
  extended objects \citep{Blanton2011,Geha2012}. This catalogue only
  include galaxies with redshifts $z<0.055$ and is almost complete for
  (model) $r$-band magnitudes $m_r<17.2$.  Once equivalent cuts are
  applied to our sample we find that the abundance of satellites in
  DR7 and NSA are consistent within their error bars; with a small
  upward offset of DR7 compared to NSA.  This shift is negligible for
  $M_*^{\rm pri} \geq 10^{10} M_\odot$ and less than a factor of $\sim
  2$ for fainter primaries, which is comparable to the uncertainties
  estimated by bootstrap of the samples.  Importantly, the apparent
  mass-independence of satellite counts for primaries less massive
  than $M_*^{\rm pri} = 10^{10} M_\odot$ is also found in the NSA
  sample.

In addition to providing hints about the power-law nature of the
$M_*$-$M_{200}$ relation at the low-mass end, satellite number counts
can also help constrain its slope.  Assuming that satellite and
primary galaxies follow the same $M_* \propto M_{200}^\beta$ power-law
relation and that the unevolved subhalo mass function is
self-similar, as predicted by CDM\footnote{$N(>\mu) \propto
  \mu^\alpha$ for small $\mu$, where $\mu=m_{\rm acc }^{\rm
    sub}/M_{\rm DM}^{\rm host}$ is the ratio of the dark matter masses
  of subhalo at infall time and host at $z=0$. Typically $\alpha \sim
  -0.75$ \citep{Giocoli2008, Yang2011}}, then the abundance of faint
satellites should scale roughly with $N(>\mu_*) \propto
\mu_*^{(\alpha/\beta)}$, where $\mu_* = m_*^{\rm sat}/M_*^{\rm pri}$.
Notice that this behaviour is only expected over the mass ranges where
the substructure mass function is a power law, which typically
requires $\mu \leq 0.1$ \citep[e.g. ][]{Giocoli2008}. This imposes an
upper limit $\mu_* = 0.1^\beta \sim 0.003$ on the relative mass of the
companions where we expect $N(>\mu_*) \propto \mu_*^{(\alpha/\beta)}$
to hold.

In the semianalytic model, $\alpha/\beta \sim 0.95/2.4 \sim 0.4$,
which is roughly in agreement with the slope of
the satellite mass function in Fig.~\ref{fig:mf_sam} measured for
low-mass companions $m*^{\rm sat}/M_*^{\rm pri} \leq 0.003$.  This
relation is independent of $M_*^{\rm pri}$, provided the simulation
resolves satellites differing by three or more orders of magnitude in
stellar mass with respect to $M_*^{\rm pri}$. This, in our
simulations, happens at $M_*^{\rm pri} \geq 10^{10}
M_\odot$. Interestingly, the scatter in stellar mass at fixed halo
mass, which in the model is $\sim 0.35$ dex for the low-mass objects
\citep{Guo2011_sam}, does not seem to impact the slope derived from the
simple arguments given previously.  However, this would change if the
scatter in the $M_*$-$M_{200}$ relation were strongly correlated to the
halo mass. In our case this correlation is rather mild.

We conclude from Fig.~\ref{fig:lumfunc_sam_sdss} that the good
agreement in shape, normalization and slope between SDSS primaries and
the mock catalogue strongly favours a power-law relation with a steep
slope $M_* \propto M_{200}^{2.5}$ between stellar mass and halo mass
of dwarf galaxies. This agrees with predictions from the semianalytic
model of \citet{Guo2011_sam} and from extrapolations of abundance-matching
studies \citep[e.g., ][]{Moster2010,Behroozi2010,Guo2010}.

\section{Discussion and Conclusions}
\label{sec:concl}

We study the abundance of satellites as a function of primary stellar
mass in the Sloan Digital Sky Survey. Using the SDSS/DR7 spectroscopic
sample from the NYU-VAGC catalogue we are able to extend previous
studies to significantly fainter primaries, $M_*^{\rm
  pri}=[10^{7.5}$-$10^{11}] M_\odot$.  In agreement with previous
work, we find that the abundance of satellites exceeding a given
satellite-to-primary stellar mass ratio, $m_*^{\rm sat} /M_*^{\rm
  pri}$, depends strongly on $M_*^{\rm pri}$ for bright primaries.  On
the other hand, {\it the abundance of satellites around dwarf
  primaries, $M_*^{\rm pri} < 10^{10} M_\odot$, is approximately
  independent of primary stellar mass}.

These results are in excellent agreement with predictions of
semi-analytic models within $\Lambda$CDM. These trends arise from the
mass invariance of substructure in CDM halos and from the varying
efficiency of galaxy formation as a function of halo mass. On dwarf
galaxy scales, where the relation between galaxy mass and halo mass is
well approximated by a steep power law, the invariance of satellite
abundance with primary mass reflects directly the scale-free nature of
substructure. Around bright galaxies the scaling between galaxy mass
and halo mass deviates from a simple power law, leading to the
observed strong increase of satellite abundance with increasing
primary mass.

 Some caveats on the statements above need to be mentioned.  The
  first relates to the definition of halo mass for a galaxy that has
  entered the virial radius of a larger object.  The mass in subhalos
  is in general ill-defined, depending not only on the identification
  algorithm \citep{Knebe2011}, but also on time, due to tidal stripping
  \citep[e.g. ][]{Tormen1998, Klypin1999b, Hayashi2003}. The stellar
  mass, however, is more resilient to tidal effects and remains
  approximately constant after accretion onto the host
  \citep{White_Rees1978, Sales2007a, Penarrubia2008}.  Thus for
  satellite galaxies, the virial mass at the moment of accretion is
  more closely related to the stellar mass than their present-day dark
  halo mass.  Since the abundance of subhaloes according to their
  infall mass --termed the ``unevolved'' subhalo mass function-- is
  also independent of host halo mass when written as a function of the
  relative mass between satellite and host $m_{\rm acc}^{\rm
    sub}/M_{\rm DM}^{\rm host}$ \citep[e.g. ][]{Giocoli2008,
    Yang2011}, the arguments given above remain valid.

  The second caveat involves the possible dependence of the
  $M_*$-$M_{200}$ relation on redshift and, related to that, whether
  satellites and primaries follow the same relation between stellar
  mass and halo mass.  Abundance matching models suggest that the link
  between the stellar mass and halo mass evolves weakly with redshift 
  \citep{Shankar2006,Conroy2009,Wang_Jing2010,Leauthaud2011,Yang2012,Moster2012,Behroozi2012}. Given
  that surviving satellites are preferentially accreted at late times
  \citep[e.g. ][]{Gao2004,Sales2007a} the $M_*$-$M_{200}$ relation for
  satellites and centrals are expected to be similar. In particular,
  self-consistent semi-analytical modelling of galaxies shows only
  small differences between the two \citep[e.g. see Fig. 9 of
  ][]{Guo2011_sam}.

  Notice that these arguments do not allow for stripping of the stars
  from satellites. Arguably, this could complicate the evolution. We
  note, however, that numerical models for dwarf galaxies suggest that
  once stellar stripping sets in, total disruption soon follows
  \citep{Penarrubia2008}.  We therefore expect partial stripping of
  stars to have minor effects in large statistical samples. On the
  other hand, several studies have suggested that accounting for total
  disruption of satellites is needed to match observations
  \citep[e.g. ][]{Weinmann2006, Kimm2009}.  We address this by
  comparing observational results with a semi-analytic model that
  explicitly treats satellite disruption by tidal forces.

Under the assumption of a $\Lambda$CDM universe, the good agreement in
shape and normalization between satellite counts in SDSS and those in
the mock catalogue provides support for a steep stellar-halo mass
relation for dwarfs, consistent with the $M_* \propto M_{200}^{2.5}$
predicted both by semi-analytic models and by extrapolations of
current abundance-matching analyses.  More definitive constraints on
the slope of the $M_*$-$M_{200}$ relation for dwarf galaxies may come
from a robust determination of the slope of satellite abundances
around isolated primaries in tandem with studies of the effect of
scatter in the stellar mass - halo mass relation. Probing increasingly
fainter companions in observational surveys of the surroundings of
isolated dwarfs may prove crucial for this goal.

\section*{Acknowledgements}
\label{acknowledgements}

The authors are grateful for a prompt and very useful referee report
that helped to improve the previous version of the manuscript.  We
thank Risa Wechsler for stimulating discussions.  LVS is grateful for
financial support from the CosmoComp/Marie Curie network. The authors
thank the hospitality of the Kavli Institute for Theoretical Physics,
Santa Barbara during the program ``First Galaxies and Faint Dwarfs:
Clues to the Small Scale Structure of Cold Dark Matter'', where part
of this work was completed.  This research was supported in part by
the National Science Foundation under Grant No. NSF PHY11-25915.

\bibliography{master}

\end{document}